# Understanding the roles of electronic effect in CO on Pt-Sn alloy surface via band structure measurements.

Jongkeun Jung[a,b], Sungwoo Kang[c], Laurent Nicolaï[d], Jisook Hong[e], Jan Minár[d], Inkyung Song[a,b], Wonshik Kyung[a,b], Soohyun Cho[f], Beomseo Kim[a,b], Jonathan D. Denlinger[g], Francisco J. C. S. Aires[h,i], Eric Ehret[h], Philip N. Ross[j], Jihoon Shim[k], Slavomir Nemšák[g] Doyoung Noh[l,m] Seungwu Han[c], Changyoung Kim[a,b,*], Bongjin S. Mun[l,m,**]

[a]Center for Correlated Electron Systems, Institute for Basic Science, Seoul 08826, Republic of Korea.
[b]Department of Physics and Astronomy, Seoul National University, Seoul 08826, Republic of Korea.
[c]Department of Materials Science and Engineering, Seoul National University, Seoul 08826, Republic of Korea.
[d]University of West Bohemia, New Technologies Research Centre (NTC), Univerzitni 8/2732, 301 00 Plzen Czech Republic.
[e]The Molecular Foundry, Lawrence Berkeley National Laboratory, Berkeley, California, 94720, USA.
[f]Shanghai Institute of Microsystem and Information Technology (SIMIT), Chinese Academy of Sciences, Shanghai 200050, People's Republic of China.
[g]Advanced Light Source, Lawrence Berkeley National Laboratory, Berkeley, California, 94720, USA.
[h]Université de Lyon, Université Claude Bernard Lyon 1, CNRS - UMR 5256, IRCELYON 2, Avenue Albert Einstein, 69626 Villeurbanne Cedex, France.
[i]Laboratory for Catalytic Research, National Research Tomsk State University, 36 Lenin Avenue, Tomsk 634050, Russian Federation
[j]Materials Science Division, Lawrence Berkeley National Laboratory, Berkeley, California, 94720, USA.
[k]Department of Chemistry, Pohang University of Science and Technology, Pohang 37673, Korea.
[l]Department of Physics and Photon Science, Gwangju Institute of Science and Technology, Gwangju, Republic of Korea.
[m]Center for Advanced X-ray Science, Gwangju Institute of Science and Technology, Gwangju, Republic of Korea.

Corresponding Author: [*]changyoung@snu.ac.kr, [**]bsmun@gist.ac.kr

Abstract: Using angle-resolved photoemission spectroscopy, we show the direct evidence of charge transfer between adsorbed molecules and metal substrate, i.e. chemisorption of CO on Pt(111) and Pt-Sn/Pt(111) 2x2 surfaces. The observed band structure shows a unique signature of charge transfer as CO atoms are adsorbed, revealing the roles of specific orbital characters participating in the chemisorption process. As the coverage of CO increases, the degree of charge transfer between CO and Pt shows clear difference to that of Pt-Sn. With comparison to DFT calculation results, the observed distinct features in the band structure are interpreted as back-donation bonding states of Pt molecular orbital to the $2\pi$ orbital of CO. Furthermore, the change in the surface charge concentration, measured from the Fermi surface area, shows Pt surface has a larger charge concentration change than Pt-Sn surface upon CO adsorption. The difference in the charge concentration change between Pt and Pt-Sn surfaces reflects the degree of electronic effects during CO adsorption on Pt-Sn

## 1. Introduction

In heterogeneous catalysis, the knowledge of surface electronic structure provides critical insight for understanding fundamental surface chemical reactivity of a material [1,2]. Conversely, desired chemical properties could be realized via modifying electronic structures of materials. In surface catalysis, the latter approach has been the major driving force for finding ideal surface catalyst. Based on the information on the *d*-band structure and chemical properties, Nørskov *et al*. suggested that the anticipated surface chemical properties could be obtained by tuning the electronic structures of alloys with optimum elements [3,4]. In this so-called *d*- band model, the structure of the $d$-band near the $E_F$ (Fermi level) displays strong correlation with the surface chemical reaction. Often, the $d$-band model has been referred to choosing the effective alloying components to maximize the synergistic effects between participating alloy elements in heterogeneous catalysts [5,6].

As a representative study of bimetallic alloy catalyst, Pt-Sn alloy has been thoroughly studied. Using XPS and DFT calculation, Rodriguez *et al.* pointed out that the presence of charge transfer from Sn to Pt as well as the following rehybridization between Pt and Sn generate unique surface chemistry of SO2, suggesting that both ensemble and electronic effects are important in the reaction [7]. In an STM study, Koel *et al.* explored the sign of charge modulation near the $E_F$ of Pt-Sn alloy surface, *e.g.*, the reduction of electron density near the $E_F$ at the Pt site, and interpreted the electronic structure modulation as a measure of chemical reactivity [8]. On the theoretical side, Sautet *et al.* calculated the amount of charge transfer from Sn to Pt depending on the surface structure of Sn [9]. It was pointed out in the work that the $d$-band center is shifted due to the charge rearrangement, highlighting the role of electronic structure due to the amount of neighbor Sn atoms.

While the major aspect of the electronic structure change has been discussed, there is no decisive experimental evidence for the modified Pt electronic structure due to Sn alloying and its participation on surface chemical reactions. The detailed information on the orbital dependent electronic structure of Pt-Sn and charge transfer in the chemical reaction can provide significant insight into the nature of reactivity of Pt-Sn alloys. It in turn can be beneficial for the development of advanced alloy catalysts. Nonetheless, information on such orbital dependent electronic structure and its effect on the catalytic reaction has not been obtained due to the lack of a suitable tool to investigate the issue.

We resolve the issue by studying ordered CO molecules on Pt(111) and Pt-Sn(111) surface alloys using angle-resolved photoemission spectroscopy (ARPES) and DFT calculation. While CO is an ideal simple molecule for model study of gas adsorption, the use of ordered surface alloy is especially important as it allows observation of clear experimental dispersions, unlike in the case of bulk alloys. With DFT calculation, the enhanced spectral features near $E_F$ are identified as $d_{xz}$ and $d_{yz}$ orbitals of the π-bonding. The charge transfer mechanism deduced from the band structure near $E_F$ is found to be described by a combination of the π- and σ-bonding interactions in the Blyholder model [10]. The inspection of band structure during chemical reaction reveals the key role of electronic effects in Pt-Sn alloy. This method can be applied to studies of other metal-based catalyst systems.

## 2. Experimental and Computational Methods

2.1 Experimental Methods.

Clean Pt(111) surfaces were prepared by repeating Ar sputtering and annealing at 1,000 K. Pt-Sn/Pt(111) 2 × 2 surface alloy was fabricated by Sn deposition and annealing on a Pt single crystal surface at 1,100 K. The surface arrangements of Pt(111), Pt-Sn/Pt(111) surface alloy and CO-adsorbed samples were monitored by LEED.(Fig. S1) ARPES measurements were performed in a laboratory-based system, equipped with a Scienta DA30 analyzer. The energy resolution was 8 meV at 21.2 eV photon energy, and the angular resolution was 0.1° which corresponds to momentum resolution of 0.002$A^{-1}$. CO molecules were dosed at 100 K for both Pt and Pt-Sn/Pt(111) surfaces. ARPES and LEED were performed at 10 K for pristine and 100 K for CO-adsorbed samples. All measurements were performed in an ultra-high vacuum chamber with a base pressure less than 4 × $10^{-11}$ Torr, and all the measurements including CO dosing were performed within 8 hours after cleaning.

2.2 Computational Details.

The DFT calculations were performed with the Vienna Ab initio Simulation Package (VASP) [11]. Vdw-DF exchange−correlation functional was used to describe the electron−electron interaction [12]. This functional is known to correctly describe the order of adsorption energies of CO on Pt sites.[13] We also checked that the use of PBE functional with Hubbard U correction [14,15], which could resolve the Pt puzzle [16], leads to similar results. A 12 layers of Pt slab was used with ~10 Å of vacuum region. To calculate Pt/Sn structure, atomic structures were relaxed until atomic forces acting on each atom were reduced under 0.03 eV/Å. For bare Pt slab, the structure geometry was fully optimized. On the other hand, for the structures of supercells (CO-adsorbed structures and Sn-doped structures), only top 4 layers were allowed to move while the bottom 8 layers were fixed. A 10 × 10 × 1 k-point mesh for Brillouin zone sampling was used as a unit cell and, for supercells, the number of mesh was reduced in proportion to the surface area. The band unfolding is performed using BandUP code [17,18].

## 3. Results and Discussion

3.1 Electronic structures of Pt and Pt-Sn/Pt(111) surface alloy

Figure 1a shows the structures of Pt(111) and Pt-Sn/Pt(111) 2 × 2 surfaces. The 1 × 1 and 2 × 2 unit-cells for Pt and the Pt-Sn/Pt(111) surface alloy, respectively, are indicated by black dashed lines. In the Pt-Sn/Pt(111) 2 × 2 surface alloy, Sn atoms (shown in red) replace surface Pt atoms to form a 2 × 2 unit cell. The surface Brillouin zone is reduced for Pt-Sn/Pt(111) due to the larger 2 × 2 surface unit cell.

We first use ARPES data to identify the band structures of clean Pt(111) and Pt-Sn/Pt(111) 2 × 2 surfaces, as shown in Figs. 1b and 1c. These high-symmetry cut data show various dispersive bands with identified bands marked with black dotted lines. As mentioned above, band folding takes place due to the 2 × 2 surface reconstruction of the Pt-Sn/Pt(111) surface. This explains the greater number of bands observed for the Pt-Sn/Pt(111) surface in Fig. 1c compared to the Pt(111) surface in Fig. 1b. To distinguish the original and folded bands, guide lines are superimposed onto the Pt-Sn/Pt(111) ARPES data in Fig. 1c. Black dotted lines indicate the original band structure, while red dotted lines indicate folded bands. It is important to note that the folded bands at the Γ point originate from unfolded bands at the M point. This band folding effect is even more evident in the Fermi surface map data and will be discussed later.

While the unfolded band structure of Pt-Sn/Pt(111) along the Γ-K direction in Fig. 1c is similar to that of Pt in Fig. 1b, significant differences exist between these two sets of data, indicated by blue, yellow, and red arrows. While apparent band splitting can be seen on the Pt surface near the M point, no such feature is discernable on the Pt-Sn/Pt(111) 2 × 2 surface (blue arrow). The band indicated by the red arrow is a new band that is not observed at the Pt surface. In addition, the shape of the band on the M-K line near $E_F$ (marked by the yellow arrow) differs between the two surfaces. These differences in the band structure are due to the surface alloying of Pt and Sn.

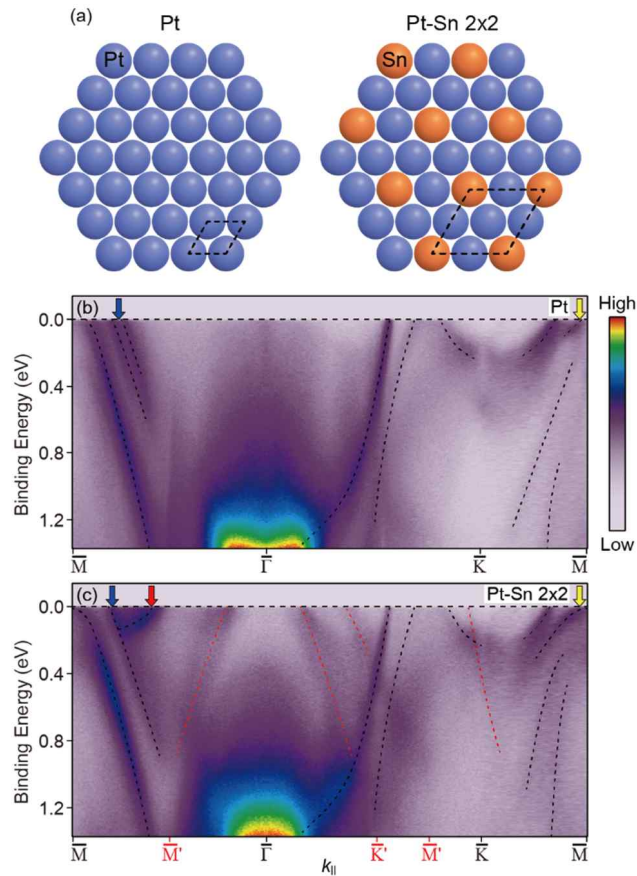

**Fig. 1.** ARPES measurements of k-space resolved electronic structure. (a) A top view shows the Pt(111) and Pt-Sn/Pt(111) 2 × 2 crystal structures. Black dashed lines represent the 1 × 1 and 2 × 2 unit-cells. The electronic structures of (b) Pt(111) surface and (c) Pt-Sn/Pt(111) 2 × 2 surface alloy are shown with guide lines. The red dotted lines represent Pt-Sn/Pt(111) 2 × 2 folded bands; unfolded bands are indicated by black dotted lines. The differences in band structure between Pt and unfolded Pt-Sn/Pt(111) 2 × 2 surface alloy are marked with blue, red and yellow arrows.

3.2 Fermi surface change and increase of conduction electron

Next, we investigate the effect of CO adsorption on the electronic structure of the metal surfaces. To better understand the correlation between band structure and adsorption, the transient surface electronic structures are monitored as CO is adsorbed onto the surfaces of Pt(111) and 2 × 2 Pt-Sn/Pt(111). Figure 2 shows the Fermi surface evolution of these surfaces upon CO dosing. Distinct differences in the Fermi surface at the Γ and M points are observed when comparing clean Pt(111) (Fig. 2a) and Pt-Sn/Pt(111) 2 × 2 (Fig. 2d). The pocket at Γ on the Pt-Sn/Pt(111) surface is due to folded bands from the M point, i.e., the folded bands at Γ in Fig. 1c, and thus looks similar to the pocket near the M point. These folded bands around Γ testify to the presence of the 2 × 2 superstructure of the Pt-Sn/Pt(111) surface. On the other hand, the hole pocket at the M point, which is also seen in Fig. 1, is larger for the Pt-Sn/Pt(111) 2 × 2 surface compared to Pt. The ARPES simulation data (Figs. S2a and b) shows details of the Fermi surface topology, and matches the experimental results shown in Figs. 2a and 2d. The presence of band folding is also reproduced in the calculations and accounts for the multiple Fermi surfaces of the Pt-Sn/Pt(111) 2 × 2 surface.

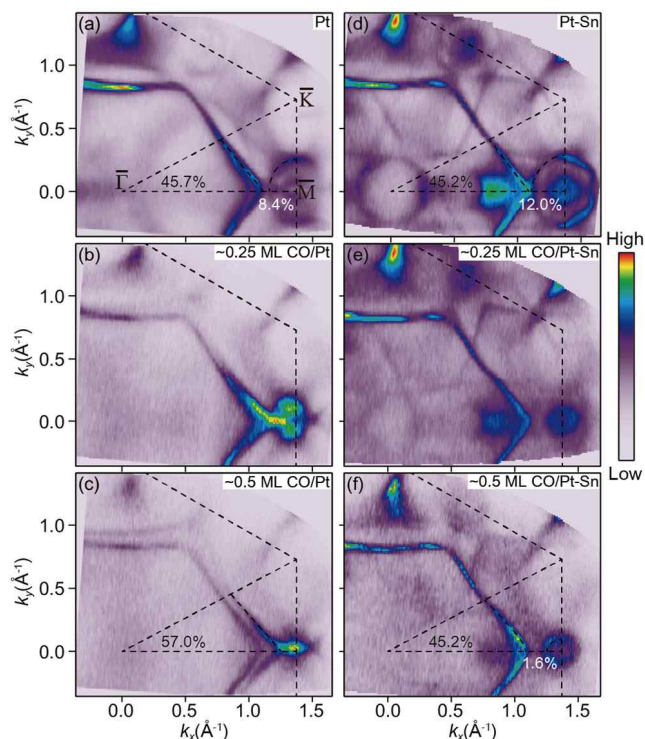

**Fig. 2**. The pristine Fermi surfaces of Pt(111) and the Pt-Sn/Pt(111) 2 × 2 surface alloy are shown along with changes in the Fermi surfaces caused by CO adsorption. Fermi surface maps are shown for (a) Pt(111), (b) ~0.25 ML CO/Pt(111), (c) ~0.5 ML CO/Pt(111), (d) Pt-Sn/Pt(111) surface alloy, (e) ~0.25 ML CO/Pt-Sn/Pt(111) surface alloy, and (f) ~0.5 ML CO/Pt-Sn/Pt(111) surface alloy. Dashed lines indicate the Brillouin zone boundary.

The data in Fig. 2 show that significant changes occur in the electronic structures of the Pt and Pt-Sn/Pt(111) surfaces as CO molecules are adsorbed. One particularly notable change upon CO adsorption is the splitting of the large hexagonal Fermi surface of Pt, centered at Γ. This Fermi surface splitting is attributed to the differential responses of surface and bulk bands of Pt to the adsorption of CO, resulting in the formation of bulk and surface states. With comparison to DFT, the splitting bands are mostly composed of in-plane orbitals, an explicit sign of surface bands (Figs. S3 and S4). Interestingly, this splitting is not observed with Pt-Sn/Pt(111), indicating the modification of the surface band of Pt upon Sn alloying, which leads to an entirely different response to CO adsorption.

More importantly, the Fermi surface pocket of Pt(111) centered at M becomes smaller as CO is adsorbed. The data in Figs. 2a-c show that the pocket at the M point of Pt almost disappears with adsorption of the ~0.5 monolayer (ML) of CO. This reduction in pocket size is less dramatic for Pt-Sn/Pt, as shown in Figs. 2d-f, with a finite Fermi pocket remaining at ~0.5 ML coverage. Since changes to the hole bands at M occur on both the Pt and Pt-Sn/Pt surfaces with CO adsorption, we examine the associated changes in electronic structure near the Fermi surfaces of both systems more closely.

First, we calculate the charge concentrations of each system. According to the Luttinger theorem, the area enclosed by the Fermi surface is proportional to the carrier number [19,20]. Thus, the number of conduction electrons of a system changed by CO adsorption can be obtained from the change in the electron and hole pocket sizes on the Fermi surface (Fig. 2) For Pt, the proportion of the large hexagonal pocket in the Brillouin zone is 45.7% and that of the M point hole pocket is 8.4%. At ~0.5 ML CO/Pt(111), the M point hole pocket vanishes and the surface hexagon pocket becomes 57% of the Brillouin zone. These changes indicate an increase in electron charge concentration of 0.394 electrons per unit cell ((57-45.7+8.4)×2=39.4 (%)). For Pt-Sn/Pt(111), the size of the hexagonal band before and after CO adsorption remained unchanged with a reduction in the size of the hole band from 12.0% to 1.6%. This indicates an increase of 0.208 electrons per 1 x 1 Pt unit cell due to CO adsorption ((12.0-1.6)×2=20.8 (%)). The change in the number of conduction electrons, thus the charge transfer, due to CO adsorption on Pt(111) is almost twice that of the Pt-Sn/Pt(111) surface alloy.

3.3 Band selective energy shift and orbital character

Next, we closely examined the band dispersion near M, which shows a significant change in the Fermi surface topology upon CO dosing. Figure 3 shows the measured electronic structures of Pt(111) and Pt-Sn/Pt(111) 2 × 2 along the M-K high symmetry cut for CO coverages of 0, 0.25, and 0.5 ML. To identify the molecular orbitals participating in charge exchange between CO and Pt, the results of DFT calculation are shown in Figs. 3d-f. In DFT calculation, the contribution from only the top surface layer is shown and the red dotted lines are drawn to help the comparison with the experimental results. In addition, the simulated electronic structures of pristine Pt(111) and Pt-Sn/Pt(111) surfaces are shown in Figs. S2c and S2d.

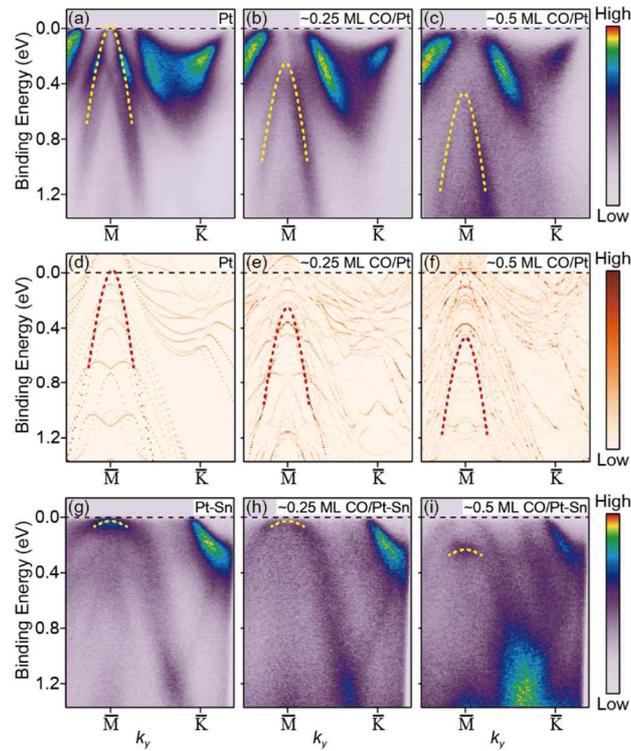

**Fig. 3.** CO-induced band shifts for Pt and the Pt-Sn/Pt(111) surface alloy. Band structures of (a) Pt(111), (b) ~0.25 ML CO/Pt(111), (c) ~0.5 ML CO/Pt(111), (g) Pt-Sn/Pt(111) surface alloy, (h) ~0.25 ML CO/Pt-Sn/Pt(111) surface alloy, and (i) ~0.5 ML CO/Pt-Sn/Pt(111) surface alloy are shown at the M point. Yellow dotted lines are provided as the guide to the eye for significant shifts induced by CO adsorbate. Theoretical (d) Pt(111), (e) ~0.25 ML CO/Pt(111), (f) ~0.5 ML CO/Pt(111) band structures are shown for comparison. Red dotted lines are duplicated from yellow dotted lines in Figs. 3(a)-(c).

In comparing the band features near M in Fig. 3a and 3(g), we note that the bands of Pt-Sn/Pt(111) are generally broader than those of Pt. The broader bands of the Pt-Sn/Pt(111) alloy, which are consistent with theoretical results (Figs. S2c S2d), arise from hybridization with Sn, and further indicate the modification of the electronic structure. Adsorption of CO results in a downward shift of the hole bands at the M point for both Pt and Pt-Sn/Pt(111) surfaces, as indicated by the yellow dotted lines. Note that this shift occurs only for the hole bands at the M point, indicating that only certain bands participate in bonding during adsorption. The band shows a rigid shift, with little or no broadening. This downshifted band is also identified in the DFT results, marked as red dotted lines in Figs. 3d-f. We found that the orbital character of the band shows an increased contribution from out-of-plane orbitals, as listed in Table 1. This finding clearly supports the π-bonding characteristics of the back-donation which occurs near $E_F$, i.e., $d_{xz}$ and $d_{yz}$ components from Pt to CO in the picture of the Blyholder model.[10]

**Table 1.** Orbital character change of energy shifting band. Calculated orbital contribution for energy shifting band of Pt during CO adsorption on Figs. 3a-c. Binding energy shows the binding energy of energy shifting band at the M point.

| Material | Binding energy[eV] | $d_{z^2}$ contribution | $d_{xz} + d_{yz}$ contribution | $d_{xy} + d_{x^2-y^2}$ contribution |
| --- | --- | --- | --- | --- |
| Pt | 0.08 | 0.592 | 0.180 | 0.107 |
| 0.33 ML CO/Pt | 0.35 | 0.388 | 0.276 | 0.076 |
| 0.5 ML CO/Pt | 0.48 | 0.297 | 0.420 | 0.038 |

While the downward shift of hole bands is similar between Pt and Pt-Sn/Pt(111), a close examination of the band behavior reveals several distinct differences. While the hole band of Pt shifts downward continuously with CO coverage, that of Pt-Sn/Pt(111) does not shift until the CO coverage reaches ~0.5 ML. This suggests that CO molecules on Pt-Sn/Pt(111) form an ordered surface only above a certain coverage as only the ordered structure is expected to contribute to the dispersive bands. This observation agrees with the result of a previous report [21], wherein the LEED pattern of a Pt-Sn/Pt(111) surface does not change with CO coverage below a certain level. In addition, the shift of the Pt hole band for 0.5 ML CO is about 500 meV (yellow line in Fig. 3c), while that for Pt-Sn/Pt(111) is only 200 meV (Fig. 3(i)). That is, at 0.5 ML of CO, the hole band shift of the Pt surface is 2.5 times larger than that of the Pt- Sn/Pt(111) surface alloy, which again reflects the difference in chemical bonding state between the two surfaces. In fact, DFT calculations show that the adsorption energy of CO on Pt(111) is larger than that of CO on Pt-Sn alloy surface. The energy difference between the two ranges from 0.16 to 0.33 eV depending on the CO coverage and calculation method [22].

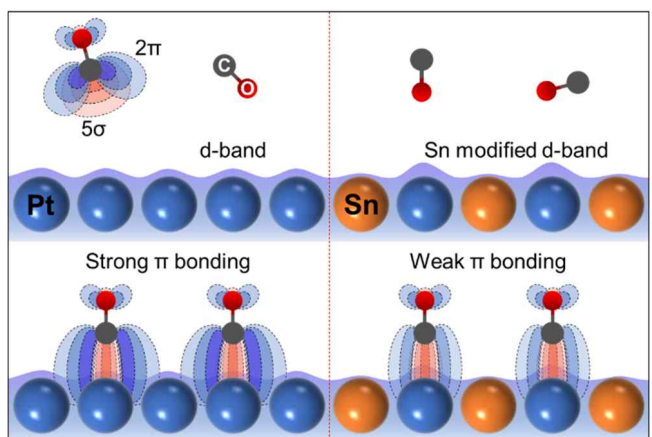

**Fig. 4.** Schematic illustration of the role of Sn on CO adsorption mechanism in Pt and Pt-Sn alloy

Here we wish to discuss the implication of our observations. The most significant change due to the Sn alloying is observed at the M point in Fig. 3. The M point band shows huge energy shift upon CO adsorption on both Pt and Pt-Sn surface alloy. Considering the dominant contribution of $d$-band orbitals to the near $E_F$ electronic structure (Fig. S3) and the increasing contribution from out-of-plane orbitals to the downward shifting hole band at M, shown in Fig. 3, the observed band downshift is a clear indicator of a π-bonding between $d$-orbitals of the metal and the 2π orbitals of CO during chemisorption of CO on the metal surface. Furthermore, the different amount of downshift of the hole band between Pt and Pt-Sn surface clearly shows that the degree of participating back-donation charge is different, i.e., weaker bonding of Pt-Sn surface, as schematically shown in Fig. 4. In fact, this is indicated in the different changes in the number of conduction electrons calculated based on the Luttinger theorem in Fig. 2. It is also to note that the increase in the conduction electrons is mainly associated with π bonding states which consist of Pt $d$-band and CO 2π states. From DFT calculations, σ-bonding states are located at high binding energies while π-bonding states exist near the Fermi level [23,24]. The lesser change in the number of conduction electrons in Pt-Sn than Pt near the Fermi surface indicates a weaker charge interaction in the case of the Pt-Sn surface, consistent with the above picture of the smaller back-donation of *d*-bands in Pt-Sn. In addition, a theoretical study of CO on Pt clusters predicted that the downward shift of Pt *d*-bands participating in the back-donation is a few hundred meV, which is consistent with our experimental observations [25]. Overall, these observations indicate that Sn plays an important role in the determination of CO adsorption properties by modifying electronic structure of Pt.

## 4. Conclusions

In summary, high resolution ARPES measurements are used to directly observe transient bonding states of CO molecules on Pt and Pt-Sn/Pt(111) alloy surfaces. Energy shifts of hole bands and changes in the Fermi pocket size near the M point of both Pt and Pt-Sn/Pt(111) show the band structure with specific orbital characters participating in the bonding process. The observed electronic band structure change indicates the role of Sn as a modifier of the Pt electronic band structure. Moreover, the amount of charge transferred from adsorbed CO is measured from the Fermi surface pockets sizes, which can be related to the different CO bonding strength between Pt and Pt-Sn surface alloy. Our ARPES results reveal the mechanism on how Sn affects the CO adsorption process on Pt, and show the important role of the surface electronic structure in molecular adsorption reactions. Our findings provide valuable information towards designing advanced catalytic materials, especially Pt-based bimetallic alloy catalyst.

**Declaration of interest**

The authors declare that they have no known competing financial interests or personal relationships that could have appeared to influence the work reported in this paper.

**Acknowledgements**

The authors thank Byungmin Sohn for helping us draw a figure in the cover letter. This work was supported by the research program of Institute for Basic Science (Grant No. IBS-R009-G2). Financial support was also provided by the National Research Foundation of Korea (NRF-2015R1A5A1009962, NRF-2017K1A3A7A09016316, NRF-2019R1A2C2008052), the GIST Research Institute Grant funded by the Gwangju Institute of Science and Technology (GIST) 2020, and the KISTI National Supercomputing Center (KSC-2020-CRE-0064). The Advanced Light Source is supported by the Office of Basic Energy Sciences of the U.S. DOE under Contract No. DE-AC02-05CH11231. This work was further supported by the European Regional Development Fund (ERDF), project CEDAMNF, reg. no. CZ.02.1.01/0.0/0.0/15_003/0000358.

**Author contributions**

**Jongkeun Jung**: Conceptualization, Methodology, Validation, Investigation, Data Curation, Writing-Original Draft, Visualization **Sungwoo Kang**: Formal analysis, Data Curation **Laurent Nicolaï**: Software, Fromal analysis **Jisook Hong**: Formal analysis **Jan Minár**: Software, Formal analysis **Inkyung Song**: Methodology, Resources **Wonshik Kyung**: Investigation **Soohyun Cho**: Investigation **Beomseo Kim**: Investigation **Jonathan D. Denlinger**: Resources **Francisco J. C. S. Aires**: Conceptualization, Resources **Eric Ehret**: Conceptualization, Resources **Philip N. Ross**: Writing-Review & Editing **Jihoon Shim**: Fromal analysis, Supervision **Slavomir Nemšák**: Investigation **Doyoung Noh**: Supervision **Seungwu Han**: Fromal analysis, Supervision, Funding Acquisition **Changyoung Kim**: Writing-Review & Editing, Supervision, Project administration, Funding acquisition **Bongjin S. Mun**: Conceptualization, Writing-Original Draft, Writing-Review & Editing, Supervision, Project administration, Funding acquisition

# Understanding the roles of electronic effect in CO on Pt-Sn alloy surface via band structure measurement.

Jongkeun Jung[a,b], Sungwoo Kang[c], Laurent Nicolaï[d], Jisook Hong[e], Jan Minár[d], Inkyung Song[a,b], Wonshik Kyung[a,b], Soohyun Cho[f], Beomseo Kim[a,b], Jonathan D. Denlinger[g], Francisco J. C. S. Aires[h,i], Eric Ehret[h], Philip N. Ross[j], Jihoon Shim[k], Slavomir Nemšák[g] Doyoung Noh[l,m] Seungwu Han[c], Changyoung Kim[a,b,*], Bongjin S. Mun[l,m,**]

[a]Center for Correlated Electron Systems, Institute for Basic Science, Seoul 08826, Republic of Korea.
[b]Department of Physics and Astronomy, Seoul National University, Seoul 08826, Republic of Korea.
[c]Department of Materials Science and Engineering, Seoul National University, Seoul 08826, Republic of Korea.
[d]University of West Bohemia, New Technologies Research Centre (NTC), Univerzitni 8/2732, 301 00 Plzen Czech Republic.
[e]The Molecular Foundry, Lawrence Berkeley National Laboratory, Berkeley, California, 94720, USA.
[f]Shanghai Institute of Microsystem and Information Technology (SIMIT), Chinese Academy of Sciences, Shanghai 200050, People's Republic of China.
[g]Advanced Light Source, Lawrence Berkeley National Laboratory, Berkeley, California, 94720, USA.
[h]Université de Lyon, Université Claude Bernard Lyon 1, CNRS - UMR 5256, IRCELYON 2, Avenue Albert Einstein, 69626 Villeurbanne Cedex, France.
[i]Laboratory for Catalytic Research, National Research Tomsk State University, 36 Lenin Avenue, Tomsk 634050, Russian Federation
[j]Materials Science Division, Lawrence Berkeley National Laboratory, Berkeley, California, 94720, USA.
[k]Department of Chemistry, Pohang University of Science and Technology, Pohang 37673, Korea.
[l]Department of Physics and Photon Science, Gwangju Institute of Science and Technology, Gwangju, Republic of Korea.
[m]Center for Advanced X-ray Science, Gwangju Institute of Science and Technology, Gwangju, Republic of Korea.

**Surface arrangement study of Pt, Pt-Sn/Pt(111) surface alloys, and CO adsorbed surfaces using LEED.**

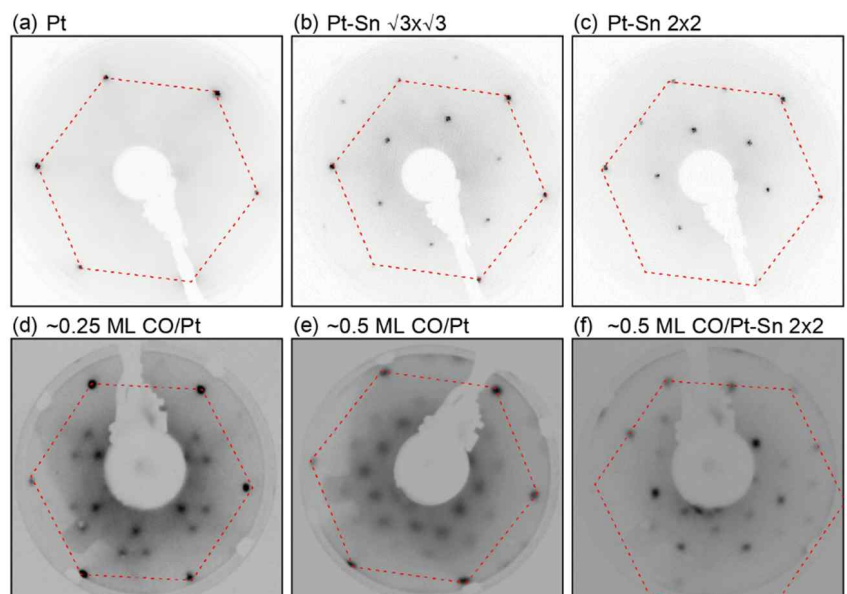

**Fig. S1.** LEED patterns of Pt(111), Pt-Sn/Pt(111) √3x√3 surface alloy, and Pt-Sn/Pt(111) 2 × 2 surface alloy, and changes in the surface arrangement by CO adsorption: (a) Pt(111), (b) Pt-Sn/Pt(111) √3x√3, (c) Pt-Sn/Pt(111) 2 × 2, (d) ~0.25 ML CO/Pt, (e) ~0.5 ML CO/Pt, and (f) ~0.5 ML CO/Pt-Sn/Pt(111) 2 × 2. Dotted lines indicate the Brillouin zone boundary.

Two different Pt-Sn/Pt(111) surface alloys, √3x√3 and 2 × 2, are prepared by Sn deposition and annealing on a Pt single crystal surface at 800 and 1,100 K, respectively. The surface arrangements of Pt(111) and the Pt-Sn/Pt(111) surface alloy are confirmed with LEED [1,2]. CO is dosed on the sample surface using gas pipe close from the sample to keep the base pressure low. While changes in LEED patterns are observed between Pt and the Pt-Sn/Pt(111) 2 × 2 surface alloy with CO adsorption, no such changes are observed with the Pt-Sn/Pt(111) √3x√3 surface; this is because CO molecules do not form an well-ordered arrangement on the √3x√3 surface due to the increased proportion of Sn [3]. CO adsorbates form ordered adlayers on Pt and Pt-Sn/Pt(111) 2 × 2 surface alloy. Two different adlayer configurations for CO/Pt and CO/Pt-Sn/Pt(111) 2 × 2 surface alloy are confirmed by comparing LEED patterns [4,5]. The real space model for CO/Pt(111) and CO/Pt-Sn/Pt(111) 2 x 2 surface alloy are provided in the references [3,5].

**ARPES simulation for Pt and Pt-Sn/Pt(111) 2 × 2 surface alloy.**

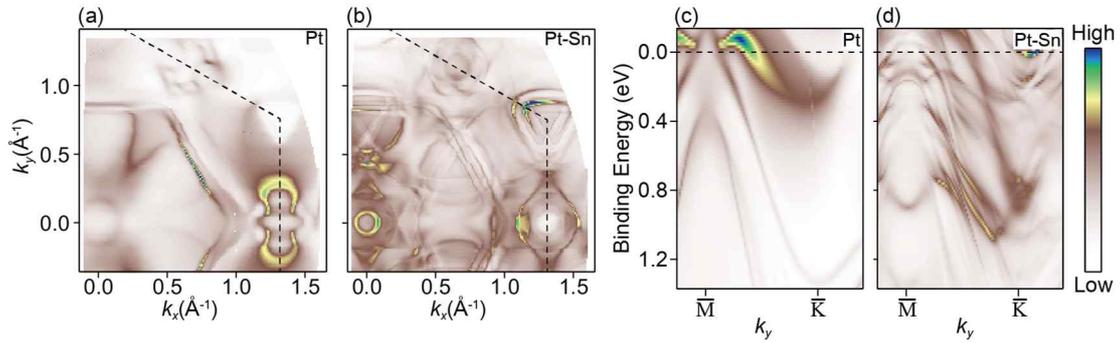

**Fig. S2.** (a) Pt, and (b) Pt-Sn/Pt(111) 2 × 2 surface alloy Fermi surfaces and band structures of (c) Pt, and (d) Pt-Sn/Pt(111) 2 × 2 surface alloy near the M point obtained with ARPES simulation.

The DFT calculations for ARPES simulation are based on a fully relativistic multiple scattering approach (Korringa–Kohn–Rostoker (KKR) method), as implemented in the SPR-KKR package [6]. As the first step in these theoretical investigations, we performed self-consistent ground state calculations for 2D semi-infinite Pt(111) and reconstructed Pt-St/Pt(111) 2 × 2 surfaces. The self-consistent results served as an input for the spectroscopic investigations. ARPES calculations are performed in the framework of the fully relativistic one-step model of photoemission using experimental geometry and photon energy [7]. Thus, the theory accounts for effects induced by light polarization, matrix-element effects, final state effects, and surface effects.

**Orbital character calculations for Pt(111)**

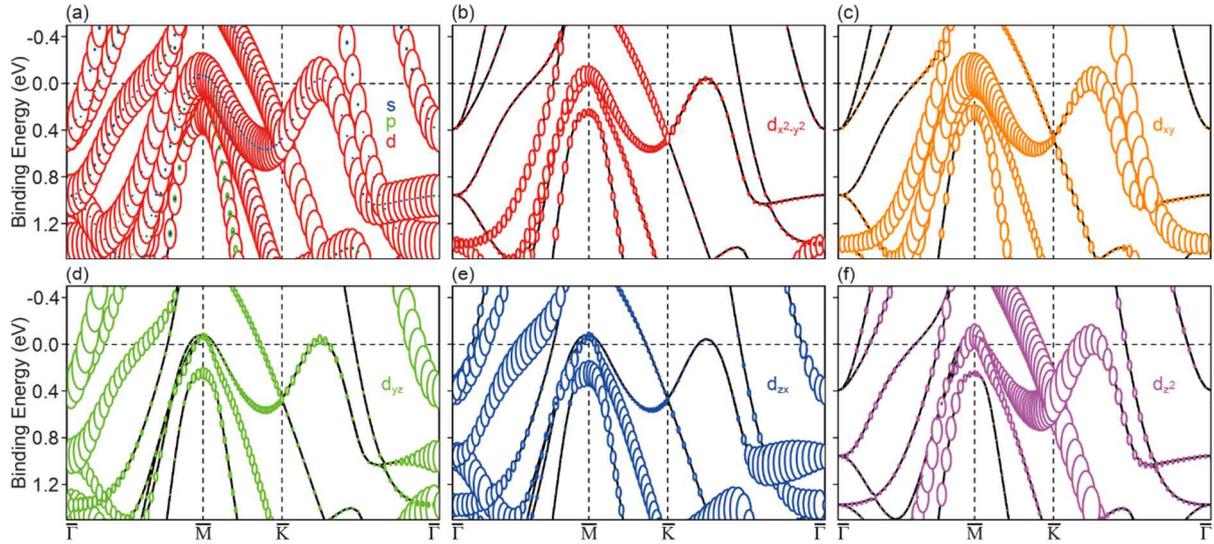

**Fig. S3.** Orbital character calculations of Pt(111). Orbital contributions of (a) s, p, d, (b) $d_{x2-y2}$, (c) $d_{xy}$, (d) $d_{yz}$, (e) $d_{zx}$, and (f) $d_{z2}$ orbitals are indicated by circles proportional to the contribution.

Figure S3 shows the orbital contribution of Pt(111) bands near the Fermi level. The orbital contribution is indicated by the size of the circle. The bands near the Fermi level of Pt(111) are mostly *d*-bands (Fig. S3a) and the detailed orbital character of these d-bands is shown in Figs. S3b-f. Figs. S3b-f shows the $d_{x2-y2}$, $d_{xy}$, $d_{yz}$, $d_{zx}$, and $d_{z2}$ orbital contributions, respectively. DFT calculations are performed with the Vienna ab initio simulation package (VASP) based on the frozen-core full-potential projector augmented-wave (PAW) method [8,9], under the generalized gradient approximation (GGA) of Perdew-Burke-Ernzerhof (PBE) as an exchange correlation functional [10]. We optimized the cell parameters of fcc Pt until the force became less than $10^{-9}$ eV/Å using a 7 × 7 × 7 Γ-centered K-points mesh. This resulted in a cell parameter of 2.81296 Å for a primitive unit cell. Based on this, we constructed a hexagonal supercell of fcc Pt featuring ABC stacking of three Pt triangular lattices. The electronic structure of the Pt supercell, which we will refer to hereafter as Pt(111), is relaxed to include spin-orbit coupling with a threshold of $10^{-3}$ eV.

**Orbital character change of hexagonal band due to the CO adsorption.**

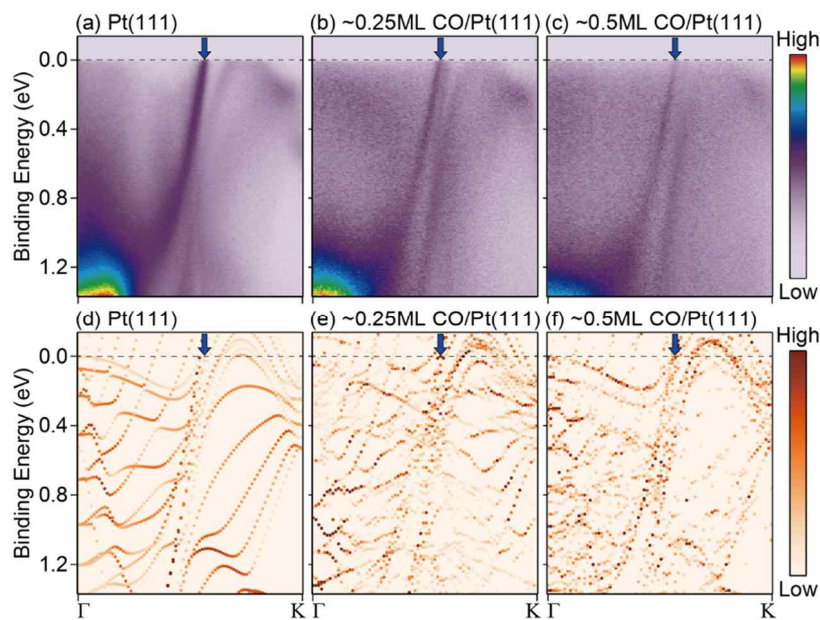

**Fig. S4.** Band structures of (a) Pt(111), (b) ~0.25 ML CO/Pt(111), and (c) ~0.5 ML CO/Pt(111) along the ΓK direction, and DFT calculation of (d) Pt(111), (e) ~0.25 ML CO/Pt(111), and (f) ~0.5 ML CO/Pt(111) along the ΓK direction. The hexagonal bands are marked with blue arrows.

Figure S4 shows the band structure of Pt(111) and CO adsorbed Pt surfaces along the ΓK direction and the DFT calculated band structure. Although it is not easy to recognize band splitting at the DFT calculation, the hexagonal bands become broader due to the CO adsorption. The orbital character calculation of the hexagonal band (Table S1) supports our intuition of surface-bulk splitting at CO/Pt(111). The Pt(111) hexagonal band is mostly made of in-plane orbital band. This in-plane orbitals have weak interaction between the layer; and cause surface-bulk splitting when CO is adsorbed on the surface. The out-of-plane orbital contribution, especially $d_{yz} + d_{xz}$ contribution of Pt(111) hexagonal band increases as CO adsorb. This orbital contribution change shows similar trend with the orbital contribution change of M point band (Table 1) and shows the electronic band structure change near Fermi level during CO adsorption comes from the back-donation π-bonding states. Calculation details for the band structure and orbital characters are written in the main text.

**Table S1.** Calculated orbital character change of hexagonal band. Calculated orbital contribution for the *hexagonal band* of Pt during CO adsorption on Figs. S4a-c.

| Material | $d_{z2}$ contribution | $d_{yz} + d_{xz}$ contribution | $d_{xy} + d_{x2-y2}$ contribution |
|---|---|---|---|
| Pt | 0.111 | 0.086 | 0.668 |
| 0.25 ML CO/Pt | 0.159 | 0.114 | 0.612 |
| 0.5 ML CO/Pt | 0.145 | 0.236 | 0.524 |

**The second derivative of CO-induced energy band shifts of Pt and Pt-Sn/Pt(111) alloy**

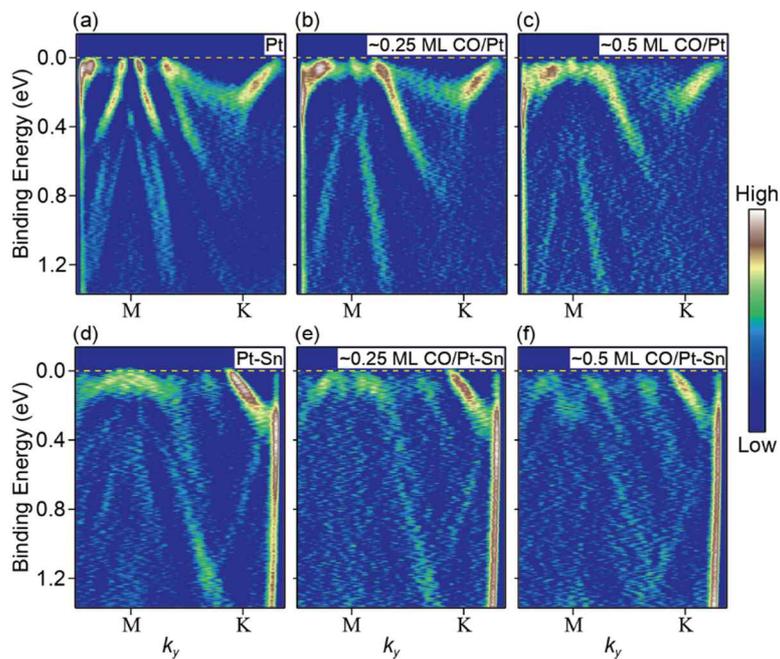

**Fig. S5.** The second derivatives of CO-induced energy band shifts are shown at the M point of (a) Pt(111), (b) ~0.25 ML CO/Pt(111), (c) ~0.5 ML CO/Pt(111), (d) Pt-Sn/Pt(111) surface alloy, (e) ~0.25 ML CO/Pt-Sn/Pt(111) surface alloy, and (f) ~0.5 ML CO/Pt-Sn/Pt(111) surface alloy.

Figure S5 shows an intensity plot of the second derivative of Fig. 3. The second derivative clearly shows the energy shifts of specific bands, indicated by yellow dotted lines in Fig. 3.